\documentclass[12pt,preprint]{aastex}

\shorttitle{Mid-IR Observations of SN1987A}
\shortauthors{Bouchet et al.}

\begin{document}

\title{High Resolution mid-Infrared Imaging of SN 1987A\footnote{Based 
on observations obtained at the Gemini
Observatory, which is operated by the Association of Universities for
Research in Astronomy (AURA), Inc. under cooperative agreement with
the NSF on behalf of the Gemini partnership: the National Science
Foundation (United States), the Particle Physics and Astronomy
research Council (United Kingdom), the National Research Council
(Canada), CONICYT (Chile), the Australian Research Council
(Australia), CNPq (Brazil), and CONICET (Argentina).}}

\author{Patrice Bouchet\altaffilmark{2,3}, James M.~De
Buizer\altaffilmark{2,4}, Nicholas B. Suntzeff\altaffilmark{2}, I. John
Danziger\altaffilmark{5}, Thomas L. Hayward\altaffilmark{4},
Charles M. Telesco\altaffilmark{6} and Christopher
Packham\altaffilmark{6}}

\altaffiltext{2}{Cerro Tololo Inter-American Observatory (CTIO),
National Optical Astronomy Observatory (NOAO), Casilla 603, La Serena,
Chile; pbouchet@ctio.noao.edu; CTIO is operated by the Association of
Universities for Research in Astronomy (AURA), Inc. under cooperative
agreement with the National Science Foundation.}
\altaffiltext{3}{ NOAO Gemini Science Center, c/o AURA, 
Casilla 603, La Serena, Chile} 
\altaffiltext{4}{Gemini Observatory, Southern
Operations Center, c/o AURA, Casilla 603, La Serena, Chile}
\altaffiltext{5}{Osservatorio Astronomico di Trieste, Via Tiepolo, 11,
Trieste, Italy} 
\altaffiltext{6}{Department of Astronomy, University of Florida,
Gainesville, FL 32611} 

\begin{abstract}

Using the Thermal-Region Camera and Spectrograph (T-ReCS) attached to
the Gemini South 8m telescope, we have detected and resolved $10 \micron$ 
emission at the position of the inner equatorial ring (ER) 
of supernova SN 1987A at day 6067. 
``Hot spots'' similar to those found in the optical and
near-IR are clearly present.  The morphology of the $10 \micron$
emission is globally similar to the morphology at other wavelengths from  
X-rays to radio. The observed mid-IR flux in the region of SN1987A
is probably dominated by emission from dust in the ER. 
We have also detected the ER at $20 \micron$ at a $4 \sigma$ level.
Assuming that thermal dust radiation is the origin of the mid-IR
emission, we derive a dust temperature of $180^{+20}_{-10} K$, and a dust
mass of $1.- 8. \times 10^{-5} M_\odot$ for the ER. 
Our observations also show a weak detection of the
central ejecta at $10 \micron$. We show that previous bolometric 
flux estimates (through day 2100) were not significantly contaminated 
by this newly discovered emission from the ER. 
If we assume that the energy input comes from radioactive decays only, 
our measurements together with the current theoretical models set a temperature of
$90 \leq T \leq 100~K$ and a mass range of $ 10^{-4} - 2. \times 10^{-3} M_\odot$ 
for the dust in the ejecta. With such dust temperatures the estimated 
thermal emission is $9(\pm3)\times10^{35}$ erg~s$^{-1}$ from the inner ring, 
and $1.5(\pm0.5)\times10^{36}$ erg~s$^{-1}$ from the ejecta.
Finally, using SN 1987A as a template, we discuss
the possible role of supernovae as major sources of dust in the Universe.

\end{abstract}


\section{Introduction}

The circumstellar envelope (CSE) surrounding SN 1987A 
consists of an inner equatorial ring (ER) flanked
by two outer rings \citep{Bur95}.  These rings, created $\sim$20,000
years before the explosion of the supernova, form the waist and caps
of an hourglass or ``bipolar" nebula $\sim$1 pc across, enclosing an
HII region and expanding into a diffuse medium that terminates in a
dense shell 4 pc in radius \citep{Cro00}.

The collision between the ejecta of SN 1987A and the ER 
predicted to occur sometime in the interval 1995-2007  
\citep{Gae97,Bor97} is now underway. ``Hot spots" have appeared inside
the ER \citep{Pun97}, and their brightness varies on time scales of a
few months \citep{Law00}.  In the next few years, new hot spots will 
continue to appear as
the whole inner rim of the ER lights up, probably brightening by a 
factor 1000 during the next several years \citep{Luo94}.

The spectral energy distribution longward of 5~$\micron$ has evolved
continuously in time.  After day 300 a cold dust-like component
appeared \citep{Sun90,Bou93,Woo93}.  An asymmetry in the
profiles of optical emission lines that appeared at day 530 showed
definitely that dust had condensed in the metal rich ejecta of the
supernova \citep{Luc91}. Although it was discovered via spectroscopy
\citep{Dan89}, the presence of the dust could be easily inferred from
the spectral energy distribution: as the dust thermalized the energy
output, after day 1000, SN 1987A radiated mainly in the mid infrared
\citep{Bou91}.  Unfortunately, after approximately day 2000, the flux
emitted from SN 1987A in the thermal spectral region was too weak to be
detected by the instruments and telescopes available.  SN
1987A was observed at day 4100 with ISOCAM onboard the ESA ISO satellite
\citep{Fis00}. \citet{Bou03} reported a weak detection of the
supernova at day 4300 with OSCIR at the CTIO 4-m telescope.  Except
for these two observations, there have been no other detections of the
mid-IR emission from the ejecta/ring region of SN 1987A over the last
five years.

There exist very few mid-infrared observations of supernovae in general.
Therefore SN 1987A, the closest known supernova in 400 years, gives us an
opportunity to explore the mid-IR properties of supernovae and their dust
with the help of the newest generation of large-aperture telescopes and
sensitive mid-IR instrumentation.

\section{Observations}

\begin{figure}[t]
\plotone{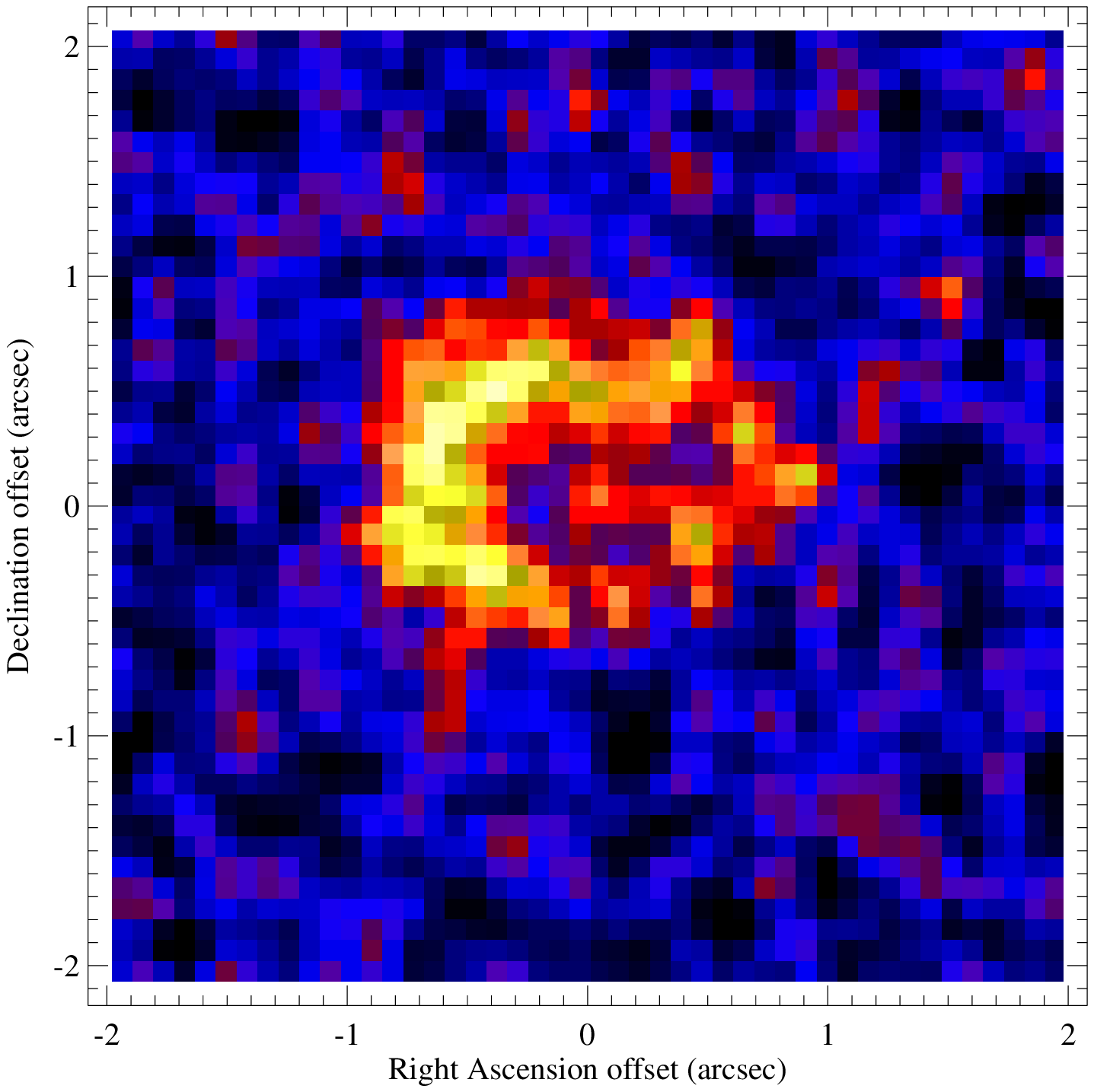}
\caption{ SN 1987A seen with T-ReCS at day 6067 in the N band filter 
($\lambda = 10.36 \micron$). The image is smoothed 2 pixels
(0.18 arcsec). Note, in particular, the central point source that corresponds 
to the ejecta of the supernova \label{fig1}}
\end{figure}

The new T-ReCS mid-IR imager/spectrometer at the Gemini-S 8m telescope
offers a combined telescope and instrument with diffraction limited
imaging (~0.3" resolution) and superbly low thermal emissivity.  On
2003 Oct 4 (day 6067), we imaged SN 1987A with T-ReCS as part of the
instrument's System Verification program.  In a 23 minute on-source
co-added image in the $N$-band filter ($\lambda$
7.70-12.97~$\micron$), we were easily able to detect and resolve the
ER (Fig. \ref{fig1}). This image shows several luminous
``hot'' spots distributed over
the ring.  The calibrated flux density integrated within an aperture
of 1.3" radius is ~9.9 $\pm$ 1.5 mJy. This absolute calibration has
been made using $\lambda_{eff}$ = 10.36 $\micron$. No color
correction was applied and this would most likely increase the flux
density.  The standard star used for this and all calibrations was HD
37160, whose flux density was taken to be 8.77 Jy in the $N$-band.
On 2003 Dec 1 (day 6125) a short exposure of the supernova in the Qa-band 
($\lambda$ 17.57-19.08~$\micron$)
led to a $3.7 \sigma$ detection of the resolved ER with a flux density of 
50.6 $\pm$ 6.6 mJy (this image will be published when reinforced by
improved S/N ratio images). We used $\alpha$ CMa with
a flux density of 44.3 Jy at 18.30~$\micron$ for the flux calibration
of that observation.

\begin{figure*}[t]
\begin{minipage}{165mm}
\plotone{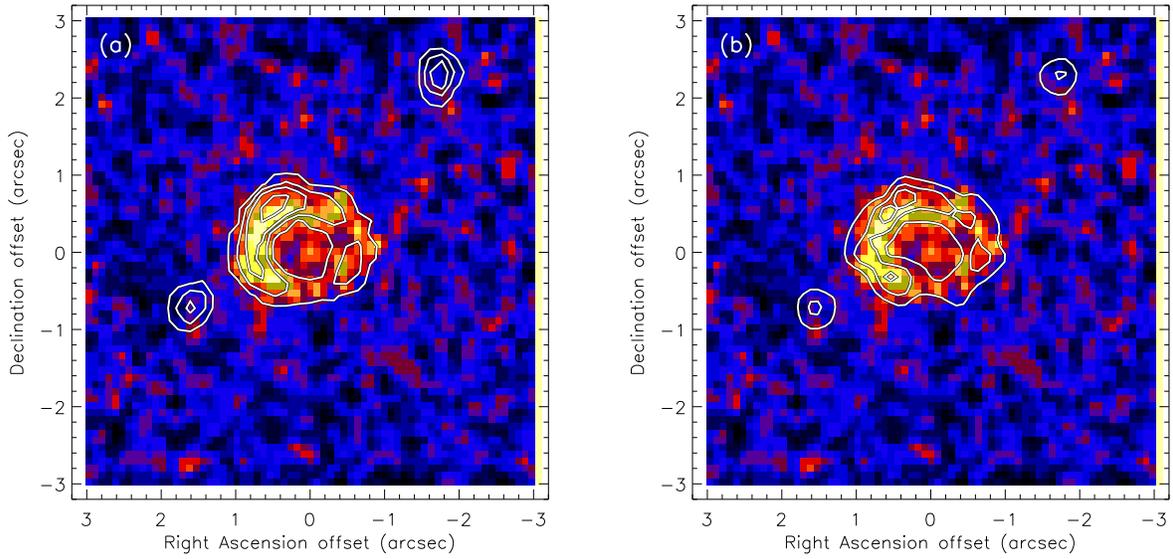}
\caption{ T-ReCS 10$\micron$ images of SN 1987A 
smoothed to 0.18 arcsec (2 pixels) resolutions 
are overlayed with contours of the images obtained in the HeI line ($1.083 \micron$)
with OSIRIS
at the CTIO Blanco telescope at day 5749 (a), and through the F656N
filter (H$\alpha$ and [N II] $\lambda 6583$ \AA) with HST at day 5555 (b). 
The HST image is smoothed
1.5 pixels (0.15 arcsec). The central source is not seen in any of
the two overlays while it is detected at $10 \micron$. \label{fig2}}
\end{minipage} 
\end{figure*}

\begin{figure*}[t]
\begin{minipage}{165mm}
\plotone{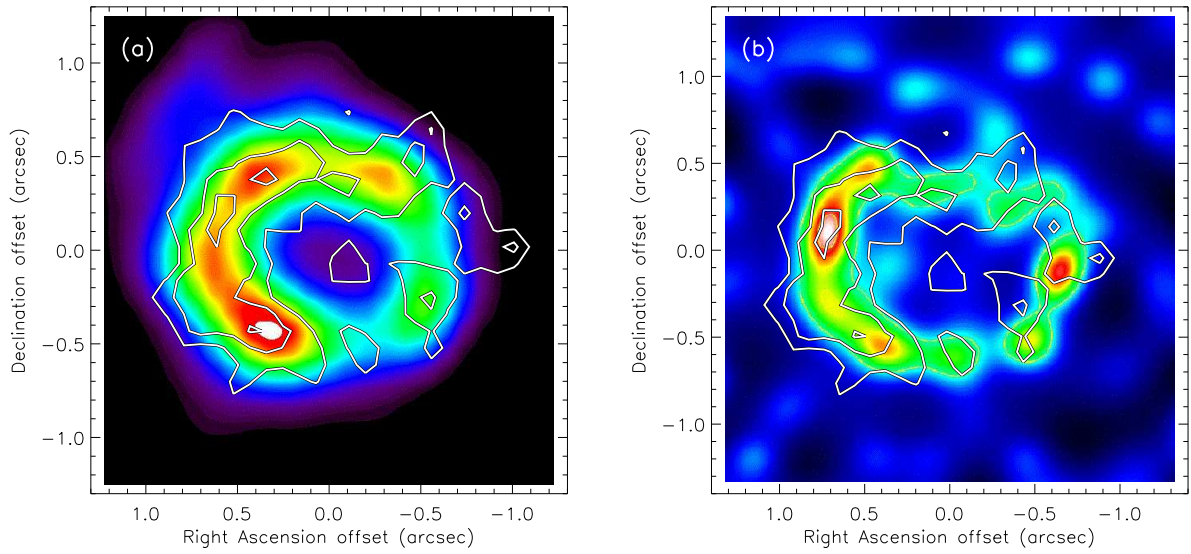}
\caption{ SN 1987A: the Chandra (a) 0.3-8.0 keV broadband image obtained at day 5791, and the ATNF
8 GHz restored image (b) obtained at day 6002.
The overlays correspond to the T-ReCS image. \label{fig3}}
\end{minipage}
\end{figure*}

In Fig.~\ref{fig2} we compare our $10 \micron$ data with data obtained
in the HeI line at CTIO \citep{Bou03} at day 5749
(Fig.~\ref{fig2}a), and in the F656N filter as observed by HST at
day 5555 (Fig.~\ref{fig2}b; 
ESO/ST-ECF Science Archive Facility). This filter includes the H$\alpha$
and [N II] ($\lambda$ 6583 \AA) lines. 
Our mid-IR image is also compared to the
{\it Chandra} Observatory X-ray image obtained at day 5791
(Fig.~\ref{fig3}a) \citep{Par04b}, and the Australian Telescope National
Facility (ATNF) 18.5 GHz (12 mm) maximum entropy restored image at day 
6002 \citep{Man03} (Fig.~\ref{fig3}b). 
There is good agreement in shape and size between our mid-IR
image and images obtained at other wavelengths. 
The mean radii and approximate surface brightness distribution
(brighter on the east side) of the ring are similar from
the X-ray to the radio, demonstrating
that the dust is co-extensive with the gas components. The origin
of that brightness asymmetry may be related either to the asymmetric
distribution of the ejecta and/or to the density variation in the
CSM \citep{Par04a,Par04b}, or to a time-dependent effect caused by 
the tilt of the ER as discussed by \citet{Pan91}.

The most likely source of mid-IR radiation is thermal emission from cool
dust (see discussion below), whereas the X-ray radiation is thermal emission
from very hot gas \citep{Par03,Par04a}. The radio emission is 
likely to be synchrotron radiation as is the case for Cas~A \citep{Dun03},
who claim that if the dust were at temperature $100-150 K$ it could
be heated by collisions with fast-moving electrons and ions in the hot
gas seen with {\it Chandra}. 
\citet{Par04a} argue that, until 2002 May 15, hard X-ray and radio 
emissions were
produced by fast shocks in the CS HII region while the optical and
soft X-ray emissions came from slower shocks in the denser ER.
\citet{Par03,Par04b} report morphological changes in their last {\it Chandra}
image, which is an indication of the blast wave now appearing to encounter 
the ER in the western side a few years after it reached the eastern side, 
and note that as of 2002 Dec. 31 (day 5791) correlations between
the X-ray and the optical/radio images are more complex than the above
simple picture.

\section{Discussion}

The origin of the 10~$\micron$ emission for both the ER and the
ejecta, may be (1) line emission from
atomic species, (2) synchrotron or free-free emission, or (3) thermal
emission from dust. As for the ER, its detection at $20 \micron$ seems
to suggest that the emission is thermal emission from dust.
A black body of temperature 
$T_{Dust} = 180^{+20}_{-10}$ can be fitted to our data. An order of
magnitude estimate can be obtained for the dust mass from the formula:
$$ M_{Dust} =
{{\pi \times D^2\times F_\nu(\lambda)}\over{\kappa(\lambda)\times \pi\times 
B_\nu(\lambda,T)}} ~~~~(1)$$
where $D$ is the distance to the supernova, $F_\nu(\lambda)$ the
observed flux at $10 \micron$, $\kappa(\lambda)$ the dust mass absorption
coefficient at this wavelength, and $\pi \times B_\nu(\lambda,T)$ the
Planck function. The approximate range for graphite and silicate grains 
for $\kappa$ at $10 \micron$ is $100-2000$ cm$^2$~g$^{-1}$. We use $D =
51.4$ Kpc \citep{Pan99}. Considering the uncertainty in the temperature we 
derive from this formula a dust mass range of $1-80 \times 10^{-6}
M_\odot$. 
\citet{Par04b} reported a plasma temperature of $T\sim2.8\times10^7$ K and a
gas density $n = 235$ cm$^{-3}$ for the hard component of their two-shock model.
According to \citet{Dwe87a} our inferred temperature indicates a high gas density 
($n\geq100$ cm$^{-3}$) and a large grain size in the ER, thus in good
agreement with the {\it Chandra} X-ray observations. 

Possible scenarios for the late-time mid-IR emission in Type II
supernovae \citep{Gra86,Ger00} are (1) dusty ejecta, (2) an infrared echo,
or (3) dust heated from circumstellar interaction.  In the last case,
the dust could be (i) pre-existing dust in the CSM heated by the outer
blast wave, or (ii) newly formed dust in the ejecta heated by a
reverse shock traveling backwards (in mass) into the supernova
envelope.  
Our observations show that the bulk of the {\it received} IR flux 
doesn't originate in the metal rich parts of the ejecta (eg. the
weak detection at the center of the ring). 
Moreover, these parts are expanding much more slowly than the outer parts
of the H-rich envelope and they have not yet reached the ring. 
As it is most unlikely
that dust formed in the tenuous outer H-rich layers of the SN interacting with
the ring, we believe that the IR emission is produced by preexisting dust in 
the CSM heated by the outer blast wave (scenario 3-i).

This interaction converts part of the kinetic energy (KE) into 
radiative output. That leads to a caveat in our reasoning:
we assume that all the emission from the ER is in the form of IR 
emission, and we do not consider the possibility that the dust may just 
be re-radiating a fraction of the ambient radiation if the heating 
mechanism were in part due to radiation and not only to collisions.
However, it is most likely that the dust may be shock-heated, in which 
case the IR emission cools
the hot plasma that gives rise to the X-ray emission as well.
Indeed, \citet{Dwe87b} show that IR emission from collisionally heated
dust is the dominant cooling mechanism of the shocked gas in SNR. 
\citet{Dwe87a} computes the ratio between the cooling function of the
gas via gas-grain collisions and that of the gas via atomic processes, 
as a function of the gas temperature only.
\citet{Dwe87b} note that, if the dust that gives
rise to the IR emission occupies the same volume as the X-ray emitting 
plasma (as appears to be the case in view of our Figure 3), this cooling
ratio translates into the infrared-to-X-ray flux ratio (IRX). These authors
show that the observed values of this ratio is significantly larger than unity for
the 9 remnants that they considered, and concluded that IR emission, mostly
attributed to gas-grain collisions, is the dominant cooling mechanism in
these remnants over large periods of their evolutionary lifetime.

\citet{Par03,Par04b} report that as the blast wave approaches the dense
circumstellar material, the contribution from the decelerated slow shock
to the observed X-ray emission is becoming significant and increasing
more rapidly than ever as of 2002 Dec. 31. They fit their data with a 
two-temperature model for that epoch, and find a temperature of
$kTe = 0.22$ keV and a luminosity $L \simeq  1.6 \times 10^{35}$ erg s$^{-1}$ 
for the decelerated slow shock component, and 
$kTe = 2.44$ keV and $L \simeq 3.7 \times 10^{35}$ erg s$^{-1}$ 
for the blast wave shock. Combining our data with
these results leads to values of $IRX \simeq 6$ and $IRX \simeq 3$ respectively.
These values fall significantly below the theoretically expected 
ratios, at roughly the same position as the Kepler remnant 
in Figure 1 of \citet{Dwe87b}. These authors argue that in Kepler
the low IRX ratio may mostly reflect the absence of dust in the
material heated by the reverse shock. Our observations then  
suggest that little dust was formed in the ejecta during the presupernova
mass loss phase of the progenitor. We note, however, that the IRX
calculations were performed for a local interstellar dust abundance and
a standard dust-to-mass ratio, which might be unapplicable for SN 1987A. 

The $N$-band detections of SN 1987A at days 4100 and 4300
suggested that dust was still present in the ejecta and the dominant
component of the ejecta's bolometric flux. However, the ISO observations
were slightly non-stellar strongly suggesting that some of the dust
emission was coming from the inner ring, presumably heated by the
shock front as the high-velocity material hits the ER material
\citep{Fis02a}.  
EUV radiation is produced in the shock, and ionizes the gas
upstream of the shock front revealing the structure and properties of
the outer nebula. 
\citet{Fis02b} estimate that the CS dust is most
likely silicate-iron or a silicate-graphite
mixture, or pure graphite. These types of grains have spectral
signatures in the $N$-band that were not detected in the dust that
condensed in the ejecta \citep{Bou90}, most probably in the form of 
X-type SiC \citep{Ama92,Luc91}. 

After day 530 the dust emission became the
dominant cooling mechanism of the ejecta of SN 1987A, radiating away the energy
from the radioactive nuclides synthesized in the explosion.
What happened then to this early dust
emission? In Fig.~\ref{fig1} a small flux enhancement with a flux
density of $0.32\pm0.1$ mJy ($3\sigma$) can be seen in the 
center of the ring. It is likely that this weak feature is the remains
of the dust emission from the condensates in the ejecta.
Although the $N$ filter contains several IR lines such as ArIII ($\lambda 
8.99 \micron$), SIV ($\lambda 10.51 \micron$), and NeII ($\lambda 12.81
\micron$), these lines require too high ionization state to contribute
significantly to this emission. Moreover, we note from Fig.~\ref{fig2}
that the ejecta has not been detected neither in the He I line, nor
in the H$\alpha$ or [N II] lines. As the ejecta has not been detected 
in the radio regime either, the possibility of free-free or synchrotron 
emission is also ruled out.
Therefore it is likely that the weak feature detected at $10 \micron$ 
is the remains of the dust emission from the condensates in the ejecta.
Probably the ER and the ejecta have different dust temperatures
as was observed in Cas~A \citep{Dun03}. We do not detect the ejecta in
our Qa-band image, which sets a flux upper limit of 41 mJy at the
3-$\sigma$ detection level at this wavelength, and then a lower limit
for the temperature of the ejecta $T\geq90$~K. 
\citet{Fra93} showed that time-dependent effects due to long
recombination and cooling times lead to a frozen-in structure of the 
ejecta of SN 1987A. However, \citet{Fra02} claim that freeze-out is
important for Hydrogen at times shorter than $\sim$3000 days, and that at 
later stages positron input from 
$^{44}$Ti and reverse shock heating are the major sources of energy
input to the envelope. In view of their Figure 2, the bolometric
luminosity of the ejecta should be of the order of $1. - 2. \times 10^{36}$ 
erg~s$^{-1}$. Assuming that there is no other energy input than
the radioactive decays, this luminosity is an upperlimit of the IR luminosity
(e.g. all the energy might not be released in the IR through dust
emission). To derive this luminosity given our 10~$\micron$ flux measurement,
the dust temperature must be $T \le 100$~K. Thus, our observations 
give us upper and lower limits, implying a temperature $90 \le T \le 100$~K 
for the dust in the ejecta.  The dust mass
derived from equation (1) is then $M = 10^{-4} - 2. \times 10^{-3}$ considering
the same approximate range for $\kappa(10\micron)$ of $100 - 2000$ 
cm$^2$~g$^{-1}$ as for the ER.

We stress that the value given here is an upper limit of the IR 
luminosity, as \citet{Dwe92}
show that at day 1153 line emission, mostly due to the [FeII] $26
\micron$ transition, and other ground-state fine structure lines at 35,
51 and 87 $\micron$, may contribute significantly to the total
luminosity. These authors find for that epoch a dust temperature 
$120 < T < 160 K$.
We note that the same caveat given for the ER applies to our statement 
concerning the energetics of the ejecta: we do not consider
the mechanical energy, which is in the process of being 
transferred to the ambient medium through the forward shock or to the
ejecta itself through the reverse shock, that can power the observed
IR emission. This issue will be tackled in a forthcoming paper.  

\begin{figure}[t]
\plotone{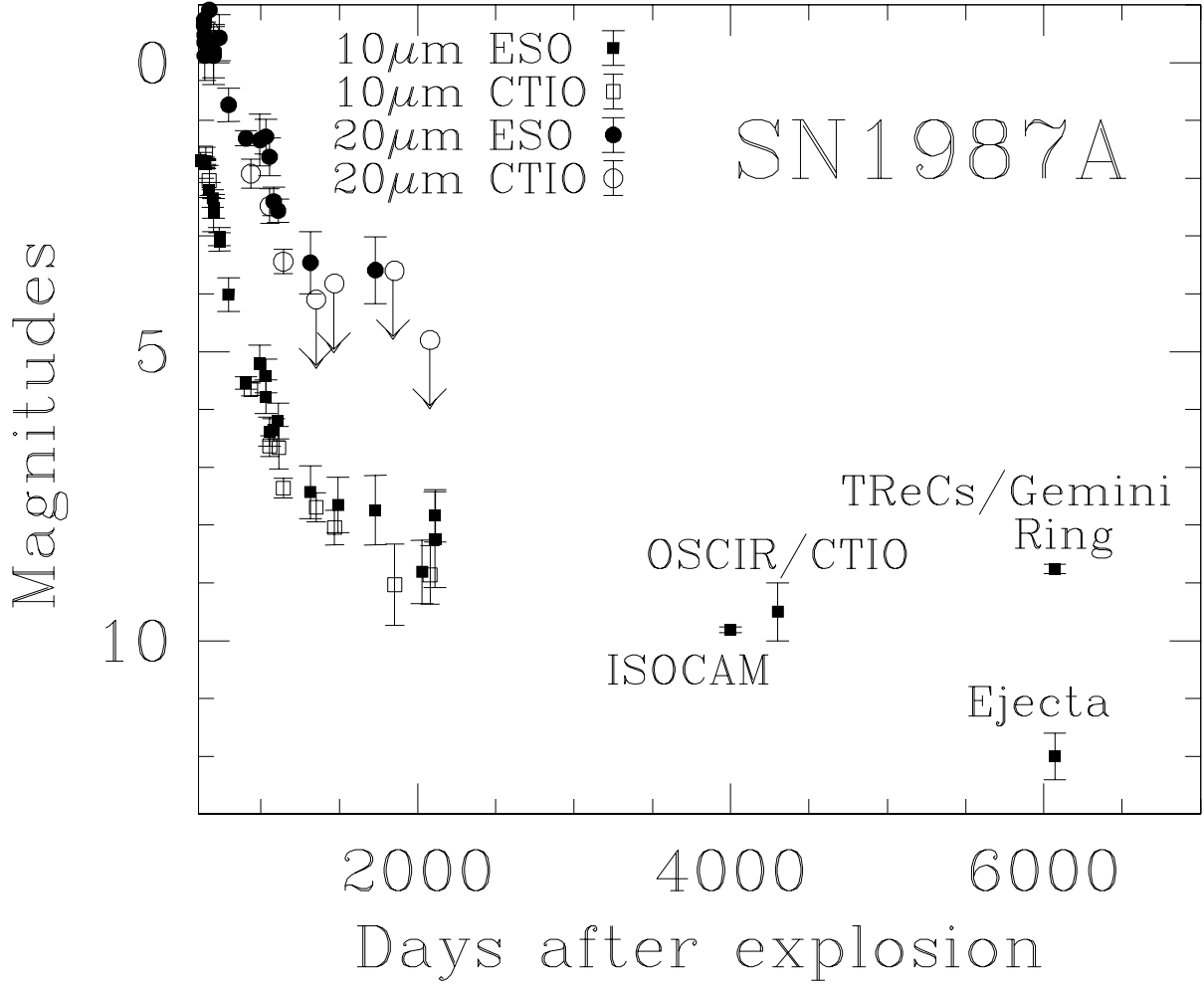}
\vspace{-4cm}
\caption{ Mid-IR light curves of SN 1987A. The point called `Ejecta'
derives from the weak point source near the center of the image \label{fig4}}
\end{figure}

Fig.~\ref{fig4} shows that the flux at 10~$\micron$ declines
exponentially from day 2200 through day 4200 (the ISOCAM and OSCIR
observations). This figure shows two data points for day 6067 (ejecta and ER).
The possible ring contribution to day 4200 is unknown. However, it
seems that the ejecta 10~$\micron$ flux from day 6067 lies on the
extension of the exponential decline curve that roughly goes through 
the 10$\micron$ flux at day 4200. If one assumes then that the 
10$\micron$ emission from the ejecta indeed follows an exponential decline law, 
that leaves no room for any ring contribution to day 4200. We conclude  
therefore that the 10~$\micron$ flux decline of the ejecta from day 2200 to 6000 is
roughly 0.32~mag~y$^{-1}$, and that the ring emission 
started most likely around day 4000 or later, in good agreement with 
Fig.~5 of \citet{Par02} for ATCA and Chandra/ROSAT data.  

\citet{Ger00} report on the detection of dust emission in the near-IR
spectra of the Type IIn supernovae SN 1998S (a supernova similar to SN
1987A) and SN 1997ab, and summarize dust emission in other supernovae:
the IR excess observed in SN 1979C and SN 1985L is interpreted
as dust forming in the ejecta, while it is attributed to dust lying in
a preexisting CSM in SN 1982E, SN 1982L, SN 1982R, SN 1993J, SN 1994Y, 
and SN 1999el \citep{Dic02}.
Although it has been claimed \citep{Dwe83a,Dwe83b} that the near-IR emission 
from SN 1980K can be explained as an IR echo, we note that this supernova 
displayed, together with an IR excess, large blueshifts of the [O I] and 
[O III] lines of $\sim$3300 km~s$^{-1}$ \citep{Fes90} compared with 
1870 km~s$^{-1}$ in 
SN 1987A, which seems to imply that dust has formed also in the ejecta
of SN 1980K. Also, \citet{Elm03} presented clear evidence for dust
formation in the Type IIP SN 1999em 500 days after explosion. 
\citet{Lag96,Dou99,Are99} detected a continuum consistent with
silicate dust of very small pyroxene
grains (MgSiO$_3$) in the youngest known Galactic SNR Cas~A.

All of these above observations of dust in supernovae may lead us to
the conclusion that supernovae could be a major 
source for interstellar dust. Various indirect arguments
\citep{Ama92} and theoretical calculations 
\citep{Dwe80,Cla82,Dwe98} also suggest that hypothesis.
Nevertheless stellar winds from various types of stars are still
considered a major source of dust production. For instance, \citet{Dou99} 
argue that supergiants and AGB stars form the bulk of the dust in the Galaxy, 
not supernovae. \citet{Dun03} argue that  
SNe are at least as important as stellar winds in producing dust in
our Galaxy and in galaxies at high redshifts.  
\citet{Mor03} also conclude that SNe, or their progenitors, may be important 
dust formation sites.

\section{Considerations for the Future}

Several theoretical models predicted the presence of dust in the CSE
of SN 1987A which was produced by a wind in the supergiant
phase. The CS dust is likely to be in the form of pyroxene or graphite
grains, while the dust condensed in the ejecta is most probably of
some type of silicon carbide.  Forthcoming imaging and spectroscopic
observations focusing on line versus continuum emission should determine
the type of emission present, and are required to clarify the issue of
the energetics of the supernova.

Finally, in order to assess the role of SNe in the production of dust in
the Universe, it is clearly important to measure
the presence of dust that survives into the formation of the remnants,
and for this, mid-IR and sub-mm observations are critical.

\acknowledgements PB expresses his warmest thanks to our referee, Eli Dwek 
for fruitful discussions and constructive comments on the original version 
of this paper. Many thanks go also to Sangwook Park for helpful
communications and for providing us with the {\it Chandra} image showed
in Figure 3a prior to publication. The contribution of Macarena Campos 
for the reduction of the OSIRIS image shown in Figure 2a is greatly
appreciated. NBS acknowledges support for the study of SN1987A
though the HST grants GO-8648 and GO-9114 for the Supernova INtensive
Survey (SInS: Robert Kirshner, PI).

\clearpage

\clearpage
\end{document}